\pdfoutput=1

\documentclass[12pt]{iopart}
\usepackage{iopams}
\usepackage{graphicx}
\newcommand{\be}{\begin{equation}}
\newcommand{\ee}{\end{equation}}
\newcommand{\eqref}[1]{(\ref{#1})}

\begin{document}

\title{Composite vortices in nonlinear circular waveguide arrays}
\author{Daniel Leykam$^1$, Boris Malomed$^2$, and Anton S. Desyatnikov$^1$}

\address{$^1$Nonlinear Physics Centre, Research School of Physics and Engineering,\\The Australian National University, Canberra ACT 0200, Australia\\
$^2$Department of Physical Electronics, School of Electrical Engineering,\\Faculty of Engineering, Tel Aviv University, Tel Aviv 69978, Israel}
\ead{daniel.leykam@anu.edu.au}

\begin{abstract}
It is known that, in continuous media, composite solitons with \textit{%
hidden vorticity}, which are built of two mutually symmetric vortical
components whose total angular momentum is zero, may be stable while their
counterparts with explicit vorticity and nonzero total angular momentum are unstable. In this work, we
demonstrate that the opposite occurs in discrete media: hidden vortex states
in relatively small ring chains become unstable with the increase of the
total power, while explicit vortices are stable, provided that the
corresponding scalar vortex state is also stable. There are also stable
mixed states, in which the components are vortices with different
topological charges. Additionally, degeneracies in families of composite vortex modes lead to the existence of long-lived breather states which can exhibit vortex charge flipping in one or both components.
\end{abstract}

\maketitle


%\noindent{\it Keywords}: discrete vortex, hidden vorticity, vector discrete nonlinear Schr\"odinger equation

\section{Introduction}

Optical vortex solitons \cite{desyatnikov_PROG}, i.e., self-trapped beams
containing phase singularities \cite%
{nye1974,soskin,Pramana,review,review2,mark}, present an ideal setting for
studying the relationship between topology and self-action effects, and may
have applications to optical data transmission and processing \cite%
{gibson2004, willner2012}. However, in local nonlinear media vortex solitons
are often destroyed by the modulational instability, and their orbital
angular momentum (OAM) is transferred to multiple filaments \cite%
{firth1997,Torner,Torner2}.

There are several approaches to suppressing the azimuthal instability of
vortex solitons. Models with competing nonlinearities, such as cubic-quintic
(CQ) \cite{Manuelo,malomed2002}, quadratic-cubic \cite{QC,Herve} and
nonlocal~media \cite{NL1,NL2} can support stable scalar vortex solitons.
Two-component vortices may also be stabilized by the CQ nonlinearity \cite%
{vectorCQ}. An alternative approach is to apply spatial
modulation to the self-defocusing nonlinearity, making its local
strength to grow, with radius $r$, faster than $r^{2}$
\cite{Barcelona}, or (more often) to apply a spatially periodic
modulation to the refractive index of a nonlinear medium
\cite{Kart}. In the limit case of the deep periodic modulation, the
medium reduces to an array of waveguides, the propagation through
which is governed by discrete equations
\cite{christo2003,lederer2008}. When the nonlinear self-action
suppresses the discrete diffraction in the arrays, discrete solitons
emerge \cite{lederer2008}. The stability of discrete vortex solitons
was predicted \cite{malomed2001,pelinovsky2005} and subsequently
observed in experiments \cite{neshev2004,fleischer2004}.
Interestingly, the stability hierarchy is inverted with respect to
continuous media: higher-order vortices, including ``supervortices"
(ring patterns built of compact discrete vortex solitons)
\cite{super}, tend to be
stable, while their lower order counterparts suffer instabilities \cite%
{higherDiscrVort,law2009, terhalle2009, desyatnikov2011}.

On the other hand, vortex solitons can also be stabilized by the action of
the cross-phase modulation (XPM) between two or more mutually incoherent
components of a composite beam. In addition to the above-mentioned
two-component model with the CQ nonlinearity \cite{vectorCQ}, examples
include multipole solitons~\cite{MP01,MP02}, two- and three-component
necklace ring patterns~\cite{desyatnikov2001,boyd,dip3_OL,dip3_OC}, and
their counterparts in the Gross-Pitaevskii equation for Bose-Einstein
condensates~\cite{BEC,Brtka}. Interactions between two vortical composite
solitons were studied in the model with the saturable nonlinearity \cite%
{Moti3}. Also studied were composite modes in which one component is
vortical, while the other one is represented by a fundamental soliton \cite%
{Moti}. In particular, solitons composed of vortex beams with oppositely
rotating vortices, including symmetric counter-rotating pairs with the
\textit{hidden vorticity} (HV), whose OAM is exactly zero, were predicted to
be much more robust than their counterparts with the explicit vorticity~\cite%
{desyatnikov2001,boyd,Brtka,desyatnikov2005,desyatnikov2007}. This feature
can be demonstrated analytically in the framework of the one-dimensional
two-component system, in which a counterpart of the HV states is represented
by \textit{hidden-momentum} counter-propagating wave pairs, with equal
amplitudes and zero total momentum \cite{desyatnikov2007}. In the general
case of an arbitrary number of symmetrically interacting components, the
stability is determined by the total OAM of the composite beam~\cite%
{Moti2,math}. Vortex solitons of the HV type were recently observed in
nematic liquid crystals~\cite{izdebskaya2012}.

A specific example of composite (\textit{two-color}) solitary vortices is
provided by those in media with the quadratic ($\chi ^{(2)}$) nonlinearity.
While they are always unstable against splitting in the uniform media \cite%
{firth1997,Torner,Torner2}, it was recently demonstrated that both single-
and two-color vortices in $\chi ^{(2)}$ media can be stabilized by an
external trapping potential \cite{HS}. HV modes can be defined in terms of $%
\chi ^{(2)}$ systems too, assuming that the fundamental-frequency beam is
built of two different components, corresponding to orthogonal
polarizations, which are parametrically coupled to a single component of the
second harmonic. The HV modes are unstable in that system (without an external potential), but the addition of the competing self-defocusing cubic nonlinearity
makes them almost completely stable \cite{HS}.

This work aims to study combined effects of the above-mentioned approaches
to the stabilization of vortices, \textit{viz}., composite discrete vortex
solitons in ring-shaped nonlinear lattices. We find that nonlinear modes of
the HV type are subject to instabilities, which demonstrates that the
above-mentioned \textit{inverse} relation between the scalar and vortex
stabilities and instabilities in discrete media, in comparison with
continua, persists for composite modes as well. An additional feature of our
discrete system is the existence of \textit{mixed-charge} composite
vortices. We conclude that their stability is tied to the topological charge
of their brightest component. Further, an azimuthal modulation applied to
discrete composite vortices may continuously deform them through a family
of modes to \textit{discrete necklace solitons}. When both components have
equal total powers, this family is degenerate, and these deformations may be
realized \textit{dynamically} by perturbing a stationary mode. This can lead
to simultaneous charge flipping of both components, similar to previous
results in continua \cite{desyatnikov2007}. Additionally, due to the broken
rotational symmetry of the discrete system, charge flipping of a single
component only may occur too.

We start by introducing the model, and obtaining analytically a class of
``separable" nonlinear modes, in Section %
\ref{sec:model}. Their stability is studied in Section \ref{sec:stability}.
The dynamics of perturbed modes and vortex-charge flipping are presented in
Section \ref{sec:dynamics}. Section \ref{sec:final} concludes the paper and
discusses possible experimental realizations.

\section{The model and vortex modes}

\label{sec:model}

We consider the discrete one-dimensional model governing the propagation of
two incoherently coupled beams with amplitudes $A_{n}(z)$ and $B_{n}(z)$
through an array of nonlinear waveguides: \numparts%
\begin{eqnarray}
&&i\partial _{z}A_{n}+A_{n-1}+A_{n+1}+(|A_{n}|^{2}+|B_{n}|^{2})A_{n}=0,
\label{eqn:model} \\
&&i\partial _{z}B_{n}+B_{n-1}+B_{n+1}+(|A_{n}|^{2}+|B_{n}|^{2})B_{n}=0.
\end{eqnarray}%
\endnumparts We apply periodic boundary conditions, $n+N\equiv n$, to define
the ring-shaped arrays [cf. Ref.~\cite{Panos,cuevas2009,alvarez2011} which
considered in detail similar equations on an infinite chain]. The system
conserves the powers, $P_{A}=\sum_{n}|A_{n}|^{2},P_{B}=\sum_{n}|B_{n}|^{2}$,
and the Hamiltonian,%
\begin{equation}
H=\sum_{n}\left[ A_{n}A_{n+1}^{\ast }+A_{n}^{\ast
}A_{n+1}+B_{n}B_{n+1}^{\ast }+B_{n}^{\ast }B_{n+1}+\frac{1}{2}\left(
|A_{n}|^{2}+|B_{n}|^{2}\right) ^{2}\right] .  \label{eqn:hamiltonian}
\end{equation}%
The current flows in the two components between adjacent sites, $n$ and $n+1$%
, are $J_{n,n+1}^{(A)}=2\mathrm{Im}(A_{n}^{\ast }A_{n+1})$ and $%
J_{n,n+1}^{(B)}=2\mathrm{Im}(B_{n}^{\ast }B_{n+1})$. Discrete vortices
correspond to circulation of the currents around the ring, and are
characterized by the integer topological charges,
\begin{equation}
m_{A}=\frac{1}{2\pi }\sum_{n}\mathrm{arg}(A_{n}^{\ast }A_{n+1}),\quad m_{B}=%
\frac{1}{2\pi }\sum_{n}\mathrm{arg}(B_{n}^{\ast }B_{n+1}),
\end{equation}%
which take values within the range of $-N/2<m_{A,B}<N/2$~\cite{desyatnikov2011,ferrando2005}%
. There is no vortex in a component if its charge vanishes.

Because the Hamiltonian (\ref{eqn:hamiltonian}) implies the Manakov-like
nonlinearity, with equal coefficients of the XPM and SPM nonlinearities \cite%
{Manakov}, it possesses an additional symmetry, associated with rotations
that mix the two components, while preserving the power at each site:
\begin{equation}
R_{1}(\varphi )=\left(
\begin{array}{cc}
\cos \varphi  & \sin \varphi  \\
-\sin \varphi  & \cos \varphi
\end{array}%
\right) ,R_{2}(\theta )=\left(
\begin{array}{cc}
\cos \theta  & i\sin \theta  \\
i\sin \theta  & \cos \theta
\end{array}%
\right) ,
\end{equation}%
where the rotation matrices act on vector $(A_{n},B_{n})$. The corresponding
conserved quantity is commonly called the isospin, $S=\sum_{n}A_{n}B_{n}^{%
\ast }$, which exists in the continuum limit too \cite{math}. Similar to the
above definition of currents $J_{n,n+1}^{\left( A,B\right) }$, quantity $%
S-S^{\ast }=2\mathrm{Im}S$ is the total isospin-power flow. Since the
coupling between components is incoherent, i.e., they cannot exchange
powers, this flow must always be zero. Consequently, a global phase shift
can always be applied to one of the components to set $\mathrm{Im}S=0$.
Therefore in this case the isospin has no physical significance, but the
fact that it is conserved during the propagation will have some consequences
later.

We look for nonlinear modes as $(A_{n},B_{n})=(U_{n}e^{i\beta
_{A}z},V_{n}e^{i\beta _{B}z})$, where $\beta _{A,B}$ are the propagation
constants and $U_{n},V_{n}$ are the site amplitudes. Here we consider a
special class of solutions, similar to necklace-ring vector solitons in bulk
nonlinear media~\cite{desyatnikov2001}, with constant total intensity on
the ring,
\begin{equation}
|U_{n}|^{2}+|V_{n}|^{2}=I\quad \mbox{for all}\quad n.  \label{I}
\end{equation}%
Under this constraint, the nonlinearity $\sim I$ is factorized, and the
stationary equations for the amplitudes separate into two effectively
decoupled linear equations, \textit{viz}.,\numparts%
\begin{eqnarray}
&(I-\beta _{A})U_{n}+U_{n-1}+U_{n+1}=&0,  \label{stat} \\
&(I-\beta _{B})V_{n}+V_{n-1}+V_{n+1}=&0.
\end{eqnarray}%
\endnumparts Exploiting the linearity of (6), we can apply the discrete
Fourier transform, $(U_{n},V_{n})=\sum_{s}(a_{s},b_{s})e^{i\Theta _{s}n}$,
where $\Theta _{s}=2\pi s/N$ is the phase winding of the Fourier mode with
the discrete vortex of charge $s$. We thus obtain an analog of the
dispersion relations~\cite{desyatnikov2011}, $\beta _{A,B}=I+2\cos \Theta
_{p,q}$, where the mode indices, $p$ and $q$, may be different for two
components with unequal propagation constants, $\beta _{A}\neq \beta _{B}$.
Note that they are degenerate with respect to the sign of the mode indices $p$ and $q$,
thus we should take the superpositions, $U_{n}=a_{+}e^{i\Theta
_{p}n}+a_{-}e^{-i\Theta _{p}n}$ and $V_{n}=b_{+}e^{i\Theta
_{p}n}+b_{-}e^{-i\Theta _{p}n}$, as a general solution. We can use the $U(1)$
invariance of each component to set $\mathrm{arg}a_{+}=-\mathrm{arg}a_{-}$, $%
\mathrm{arg}b_{+}=-\mathrm{arg}b_{-}$ without the loss of generality.
Applying the change of variables, $a_{\pm }\equiv r_{\pm }e^{\pm i\chi _{1}}$%
, $b_{\pm }\equiv s_{\pm }e^{\pm i\chi _{2}}$, and $\Delta \equiv
(1/2)(I-r_{+}^{2}-r_{-}^{2}-s_{+}^{2}-s_{-}^{2})$, the constraint equation~(%
\ref{I}) takes the form of
\begin{equation}
\Delta =r_{+}r_{-}\cos [2(\Theta _{p}n+\chi _{1})]+s_{+}s_{-}\cos [2(\Theta
_{q}n+\chi _{2})],  \label{eqn:constraint}
\end{equation}%
which must be satisfied for each value of $n=1,...,N$. This will fix some of
the six parameters $r_{\pm },s_{\pm },\chi _{1,2}$. Solutions with
symmetries will have redundancies in the $N$ constraint equations, leaving
free parameters. In the continuum limit, there is an infinite number of
constraints, hence only symmetric solutions survive in that limit. On the
other hand, additional modes may exist in the discrete system with a small
number of sites.

\begin{figure}[tbp]
\center
\includegraphics[width=80mm]{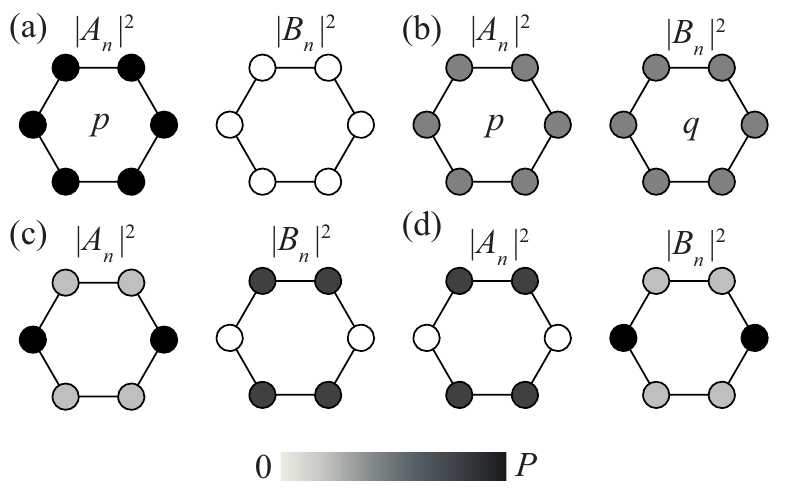}
\caption{Examples of different modes belonging to family (8)-(10) for the
ring with $N=6$. (a) A scalar vortex ($\protect\phi =\protect\varphi =0$)
with charge $p$. (b) A symmetric vector vortex ($\protect\phi =\frac{\protect%
\pi }{4},\protect\varphi =0$) of charge ($p,q$). (c) A discrete necklace ($%
\protect\phi =\frac{\protect\pi }{4},\protect\varphi =\frac{\protect\pi }{4}$%
). (d) Another discrete necklace, with $\protect\phi =\frac{\protect\pi }{4},%
\protect\varphi =-\frac{\protect\pi }{4}$.}
\label{fig:system}
\end{figure}

We will focus on a particular family of solutions with three parameters $%
(P,\phi ,\varphi )$,
\begin{eqnarray}
U_{n} &=&\sqrt{P}\left( \cos \phi \right) [\left( \cos \varphi \right)
e^{i\Theta _{p}n}+\left( \sin \varphi \right) e^{-i\Theta _{p}n}],  \label{U}
\\
V_{n} &=&\sqrt{P}\left( \sin \phi \right) [\left( \cos \gamma \right)
e^{i\Theta _{q}n}-\left( \sin \gamma \right) e^{-i\Theta _{q}n}],  \label{V}
\\
\gamma  &=&\frac{1}{2}\sin ^{-1}\left( \left( \cot ^{2}\phi \right) \sin
\left( 2\varphi \right) \right) ,  \label{gam}
\end{eqnarray}%
which can be seen as a discrete generalization of the necklace-ring vector
solitons~\cite{desyatnikov2001}. Here $P$ sets the total power of the mode, $%
\phi \in \lbrack 0,\pi /2]$ determines the relative power in its two
components, and $\varphi \in \lbrack -\pi /4,+\pi /4]$ defines the azimuthal
modulation. Notice that when the two components have different powers ($\phi
\neq \pi /4$), they require different modulation strengths, $\gamma \neq
\varphi $, to maintain the constant intensity constraint (\ref{I}).
Obviously, (\ref{gam}) has a real solution only if $|\varphi |\leq
(1/2)\sin ^{-1}(\tan ^{2}\phi )$. At the maximum allowed value of $|\varphi |
$, the intensity of the second component is zero at some sites. By varying $%
\phi $ and $\varphi $, a scalar vortex ($\phi =0,\varphi =0$) can be
continuously deformed into a \textit{symmetric composite vortex} ($\phi =\pi
/4,\varphi =0$) or a \textit{discrete necklace soliton} ($\phi =\pi
/4,\varphi =\pi /4$). Examples of these types of the modes are displayed in
figure \ref{fig:system}.

In addition to solutions with the \textit{explicit} ($p=q$) and \textit{%
hidden} ($p=-q$) vorticities, for even $N$ there also exist solutions with
mixed vorticity, $p=\pm (N/2-q)$. These appear because the constraint (\ref%
{eqn:constraint}) only needs to be satisfied at a discrete set of points. %

\section{The linear stability}

\label{sec:stability}

Linear stability is studied by introducing small perturbations of the form $%
A_{n}=(U_{n}+a_{n}e^{\lambda z}+b_{n}^{\ast }e^{\lambda ^{\ast }z})e^{i\beta
_{A}z}$, $B_{n}=(V_{n}+c_{n}e^{\lambda z}+d_{n}^{\ast }e^{\lambda ^{\ast
}z})e^{i\beta _{B}z}$ and linearizing (1) to derive an eigenvalue
problem for $\lambda $. It is convenient to introduce vectors $\mathbf{v}%
_{n}=(a_{n},b_{n},c_{n},d_{n})$, $\mathbf{E}_{n}=(A_{n},A_{n}^{\ast
},B_{n},B_{n}^{\ast })$, writing the eigenvalue problem as
\begin{equation}
(H_{L}+H_{NL})\mathbf{v}_{n}=i\lambda \mathrm{diag}(1,-1,1,-1)\mathbf{v}_{n},
\label{eqn:stability}
\end{equation}%
%
%\be
%(H_D + H_M ) \left( \begin{array}{c} a_n \\ b_n \\ c_n \\ d_n \end{array} \right) = i \lambda \left( \begin{array}{cccc} 1 & 0 & 0 & 0 \\ 0 & -1 & 0 & 0 \\ 0 & 0 & 1 & 0 \\ 0 & 0 & 0 & -1 \end{array} \right)  \left( \begin{array}{c} a_n \\ b_n \\ c_n \\ d_n \end{array} \right),
%\ee
where $H_{L}$ is diagonal in the component space (i.e., it does not couple $%
a_{n}$ to $b_{n}$, etc.), but couples adjacent sites:
\[
H_{L}a_{n}=(I-\beta _{A})a_{n}+a_{n-1}+a_{n+1},
\]%
and similarly for the other components (with $\beta _{A}$ replaced by $\beta
_{B}$ for $c_{n},d_{n}$). On the other hand, $H_{NL}$ couples different
components, rather than different sites:
\[
H_{NL}=\left(
\begin{array}{cccc}
|U_{n}|^{2} & U_{n}^{2} & U_{n}V_{n}^{\ast } & U_{n}V_{n} \\
U_{n}^{\ast 2} & |U_{n}|^{2} & U_{n}^{\ast }V_{n}^{\ast } & U_{n}^{\ast
}V_{n} \\
U_{n}^{\ast }V_{n} & U_{n}V_{n} & |V_{n}|^{2} & V_{n}^{2} \\
U_{n}^{\ast }V_{n}^{\ast } & U_{n}V_{n}^{\ast } & V_{n}^{\ast 2} &
|V_{n}|^{2}%
\end{array}%
\right) .
\]%
This can be compactly written as an outer product, $H_{NL}=\mathbf{E}%
_{n}\otimes \mathbf{E}_{n}^{\dagger }$.

\begin{figure}[tbp]
\center
\includegraphics[width=80mm]{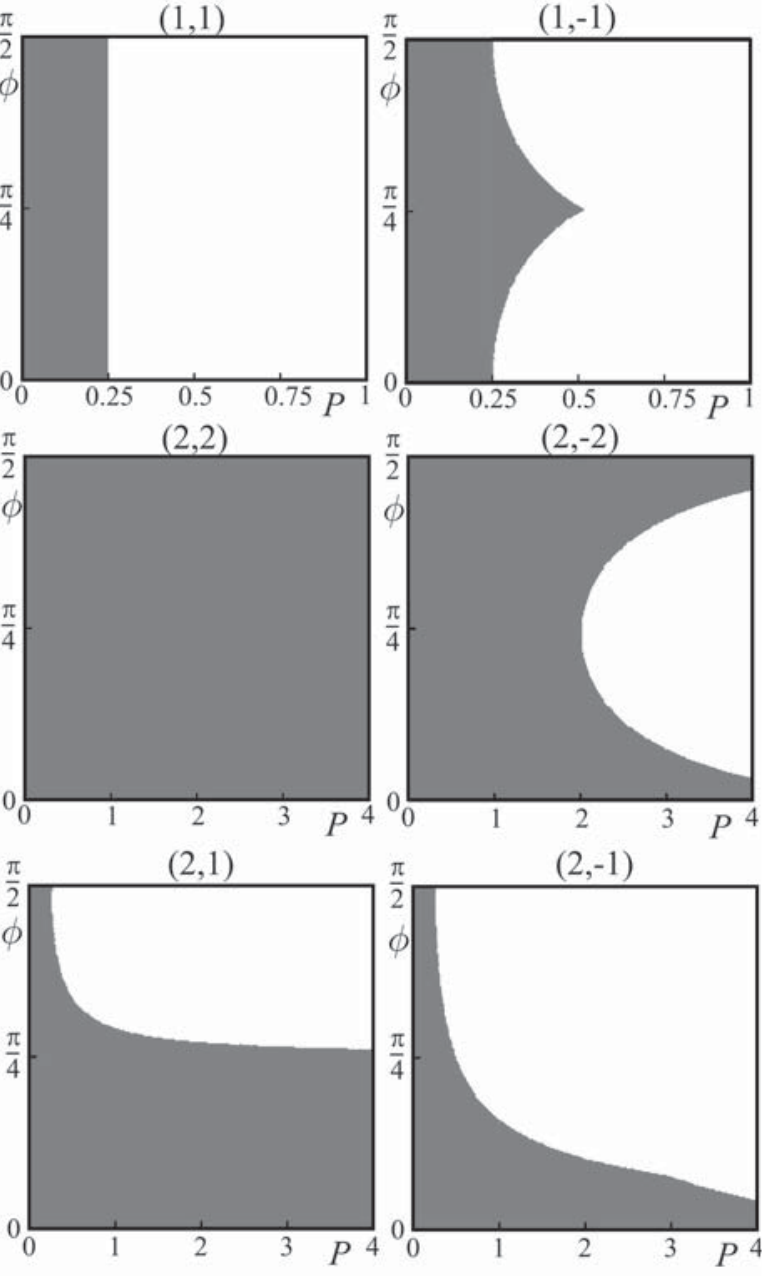}
\caption{Stable (shaded) and unstable (white) regions in the $(P,\protect%
\phi )$ parameter plane for composite vortices with different topological
charges, indicated in the brackets ($p,q$), in the ring built of $N=6$ sites. Families of hidden-vorticity modes
are represented by the top and middle panels in the right column.}
\label{fig:stability}
\end{figure}

Equations (\ref{eqn:stability}) can be solved numerically, as an
eigenvalue problem of dimension $4N$. For large $N$, it is relevant
to consider some simple limits. For example, when $\varphi =0$ the
problem simplifies significantly, the application of the discrete
Fourier transform decoupling it into a set of ``smaller" eigenvalue
problems, each of dimension 4:
%Linear stability is studied in the usual manner by introducing small perturbations of the form $E_n^{(1)} = (B_1 + u_n) e^{i \Theta_p n + i \alpha z}$, $E_n^{(2)} = (B_2 + v_n) e^{i \Theta_q n + i \beta z}$, where $B_1 = \sqrt{P} \cos \phi$ and $B_2 = \sqrt{P} \sin \phi$. We set $u_n = a_n \exp ( \lambda z ) + b_n^* \exp( \lambda^* z )$, $v_n = c_n \exp (\lambda z ) + d_n^* \exp (\lambda^* z)$ and obtain the following eigenvalue problem
%\begin{eqnarray}
% - i \lambda a_n &= a_n ( B_1^2 - 2 \cos \Theta_p ) + a_{n-1} e^{-i \Theta_p} + a_{n+1} e^{i \Theta_p } + B_1^2 b_n + B_1 B_2 (c_n + d_n ), \\
%i \lambda b_n &= b_n ( B_1^2  - 2 \cos \Theta_p ) + b_{n-1} e^{i \Theta_p} + b_{n+1} e^{-i \Theta_p } + B_1^2 a_n + B_1 B_2 (c_n + d_n ), \\
% - i \lambda c_n &= c_n ( B_2^2 - 2 \cos \Theta_q ) + c_{n-1} e^{-i \Theta_q} + c_{n+1} e^{i \Theta_q } + B_2^2 d_n + B_1 B_2 (a_n + b_n ), \\
%i \lambda d_n &= d_n ( B_2^2 - 2 \cos \Theta_q ) + d_{n-1} e^{i \Theta_q} + d_{n+1} e^{-i \Theta_q } + B_2^2 c_n + B_1 B_2 (a_n + b_n ).
%\end{eqnarray}
%Applying the discrete Fourier transform, the linear stability problem reduces to the study of the eigenvalues of the following matrix:
\[
\left(
\begin{array}{cc}
L_{A} & M \\
M & L_{B}%
\end{array}%
\right) \mathbf{v}_{s}=i\lambda \mathbf{v}_{s},
\]%
where the matrices are defined as
\begin{eqnarray}
&&L_{A}=\left(
\begin{array}{cc}
P\cos ^{2}\phi +\kappa _{p,+} & P\cos ^{2}\phi  \\
-P\cos ^{2}\phi  & -P\cos ^{2}\phi -\kappa _{p,-}%
\end{array}%
\right) , \\
&&L_{B}=\left(
\begin{array}{cc}
P\sin ^{2}\phi +\kappa _{q,+} & P\sin ^{2}\phi  \\
-P\sin ^{2}\phi  & -P\sin ^{2}\phi -\kappa _{q,-}%
\end{array}%
\right) , \\
&&M=\frac{P}{2}\sin 2\phi \left(
\begin{array}{cc}
1 & 1 \\
-1 & -1%
\end{array}%
\right) ,
\end{eqnarray}%
%\be
%M = \left( \begin{array}{cccc} B_1^2 + \kappa_{p,+} & B_1^2 & B_1 B_2 & B_1 B_2 \\
%-B_1^2 & -B_1^2 - \kappa_{p,-} & -B_1 B_2 & - B_1 B_2 \\
%B_1 B_2 & B_1 B_2 & B_2^2 + \kappa_{q,+} & B_2^2 \\
%-B_1 B_2 & -B_1 B_2 & -B_2^2 & -B_2^2 - \kappa_{q,-} \end{array} \right),
%\ee
with $\kappa _{m,\pm }\equiv 2(\cos \Theta _{m\pm s}-\cos \Theta _{m})$, and
$|s|\leq N/2$. In general, the eigenvalues cannot be obtained in an explicit
form, but simple expressions are available in some limits. Stability
diagrams of family (\ref{U})-(\ref{gam}) in the limit of $\varphi =0$ for $%
N=6$ are shown in figure \ref{fig:stability} for different combinations of
topological charges.

When both components have the same charge (which corresponds to the explicit
vorticity), the instability threshold is independent of $\phi $, and the
stability problem reduces to that for the scalar case: lower charges are
unstable, while higher ones are stable \cite{desyatnikov2011}. Another
situation takes place for the HV\ states with hidden vorticity: the HV mode
(1,-1) encounters instability at a higher power than its scalar counterpart;
ultimately, the (1,-1) mode becomes unstable at large powers, in contrast to
stable HV\ solitons in continua \cite{desyatnikov2007}. Further, the HV
configuration of the (2,-2) type also becomes unstable at large powers, in
contrast to its explicit-vorticity (2,2) counterpart, which is completely
stable.

Similar stability features are exhibited by the mixed vortex states of the
(2,1), (2,-1) types: the state with the greater total charge $p+q$ has a
larger stability region. The stability of these solitons depends on the
topological charge of the brighter component, i.e., whether the brighter
component is stable in the scalar case.

A qualitatively similar behavior is observed for other values of $N$, with
the stability dependent on whether the vortex charges are low (smaller than $%
N/4$) or high (larger than $N/4$). Vortices with the charges equal to $N/4$
represent a special case, as they form a one-parameter family of asymmetric
vortex modes, introduced in \cite{alexander2004}. This fact complicates the
stability analysis, as the stability also depends on the additional
parameter.

When $\phi =\pi /4$, an additional pair of zero eigenvalues appears for the
hidden- and mixed vortex modes $(p,-p),(p,N/2-p),(p,p-N/2)$, and the linear
stability analysis can no longer predict whether the modes are stable. The
zero eigenvalues are associated with the degeneracy of family (\ref{U})-(\ref%
{gam}) with respect to $\varphi $. Calculating the values of the conserved
quantities for the family, we obtain $P_{A}=NP\cos ^{2}\phi $, $P_{B}=NP\sin
^{2}\phi $, $H=2NP(\cos ^{2}\phi \cos \Theta _{p}+\sin ^{2}\phi \cos \Theta
_{q})+NP^{2}/2$ and $S=(1/2)NP\sin \left( 2\phi \right) [\cos (\varphi
+\gamma )\delta _{p,q}+\sin (\varphi -\gamma )\delta _{p,-q}]$.

We see that for the HV mode with ($p,-p$), when $\phi =\pi /4$, all
quantities are independent of the azimuthal modulation $\varphi $, which
means that, under small perturbations, the input with $\varphi =0$ can cycle
through solutions with different values of $\varphi $, hence we must
consider the stability of the family as $\varphi $ is varied. This family
can be obtained by applying an isospin rotation to the $\varphi =0$ mode
\cite{math}:
\begin{equation}
\left(
\begin{array}{c}
U_{n} \\
V_{n}%
\end{array}%
\right) =R_{1}(\varphi )\sqrt{P}\left(
\begin{array}{c}
e^{i\Theta _{p}n} \\
e^{-i\Theta _{p}n}%
\end{array}%
\right) ,
\end{equation}%
which follows from the fact that (\ref{gam}) is solved by $\gamma
=\varphi $ in this case. This rotation commutes with operator $H_{L}+H_{NL}$
in the linear stability problem, see (\ref{eqn:stability}). Following
the result of \cite{math}, the stability is consequently independent of $%
\varphi $. This result is supported by direct numerical solutions of the
linear stability problem. We stress that for other value of $\phi $, the
stability does depend on $\varphi $.

\begin{figure}[tbp]
\center
\includegraphics[width=80mm]{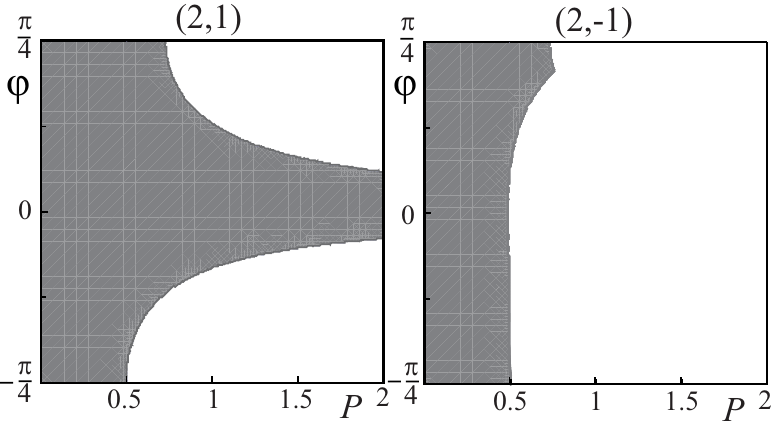}
\caption{The stability of mixed vortex modes as a function of the
azimuthal-modulation parameter ($\protect\varphi $) and total power $P$.
Shaded: marginal stability. Unshaded: linear instability}
\label{fig:stability2}
\end{figure}

The conserved quantities for the mixed vortex modes are independent of $%
\varphi $ for any value of $\phi $. However, the above reduction of the
stability problem to the $\varphi =0$ case is not relevant, as the family
cannot be generated by the isospin rotation. Therefore, the stability of the
family depends on $\varphi $, as shown in figure \ref{fig:stability2}. The
azimuthal modulation lowers the instability threshold for the explicit
vortex modes, and it can slightly increase the threshold for the HV mode. In
practice, the family will become unstable when $P$ exceeds the lowest
instability threshold that occurs as $\varphi $ is varied.

\section{Dynamics}

\label{sec:dynamics}

The existence of zero eigenvalues in the linear stability problem means that
higher-order terms will determine whether the degenerate families are
stable. To check the stability, (1) was solved numerically using
perturbed initial conditions.

\begin{figure}[tbp]
\center
\includegraphics[width=60mm]{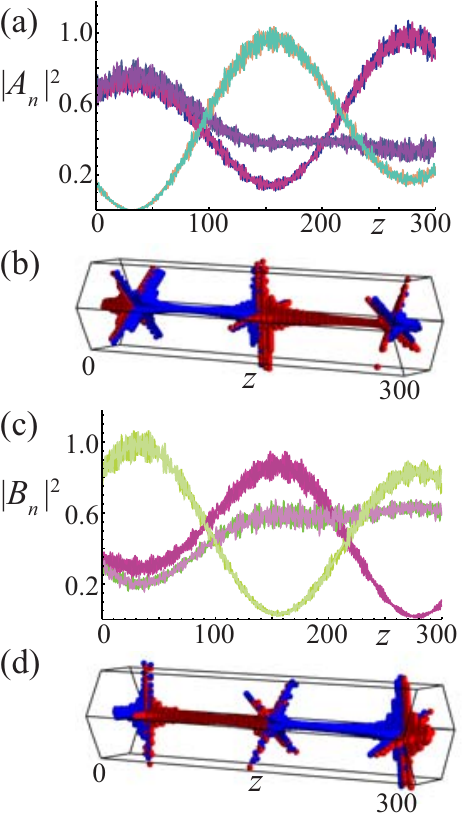}
\caption{Adiabatic dynamics of the degenerate hidden-vortex family, with
parameters $P=1,\protect\phi =\protect\pi /4$, $\protect\varphi =\protect\pi %
/8$, topological charges (2,-2), and random perturbations added to the
initial conditions at the 10\% level. (a,c): On-site powers for the two
components. (b,d): Vortex lines (red -- positive charge, blue -- negative
charge).}
\label{fig:flip1}
\end{figure}

Figure \ref{fig:flip1} shows the propagation of a perturbed charge-2 HV mode
in the waveguide ring built of six sites ($N=6$). The high-frequency
oscillations of site powers correspond to frequencies of stable eigenvalues,
while the additional low-frequency oscillations correspond to the zero
eigenvalue. During this oscillation, the power in each component acquires an
azimuthal modulation, however the sum $I_{n}=|A_{n}|^{2}+|B_{n}|^{2}$
remains (on average) constant.

When $z\approx 25$, the amplitudes of the $m=2$ and $m=-2$ Fourier modes are
equal and the topological charges of both components vanish [see figure \ref{fig:flip1}(b,d)]. Each component has two pairs of sites with equal powers, and the power vanishes at a pair of sites in one
component. The beam profiles at this point resembles those in figure \ref%
{fig:system}(c), which corresponds to a discrete necklace beam with $%
\varphi =\pi /4$. Increasing $\varphi $ further, the charges of both
components flip, and at $z\approx 100$ the azimuthal modulation vanishes, as
$\varphi $ attains value $\pi /2$. There is a second charge flip at $%
z\approx 150$ ($\varphi =3\pi /4$), this time with the other component
hosting the pair of vanishing site powers.

Thus we see that the slow oscillations correspond to an adiabatic cycling
through the degenerate family of solutions parametrized by $\varphi $. No
member of the family is subject to linear instability, therefore the
oscillations persist indefinitely long (in excess of $\Delta z=5000$,
according to our numerical results).

\begin{figure}[tbp]
\center
\includegraphics[width=60mm]{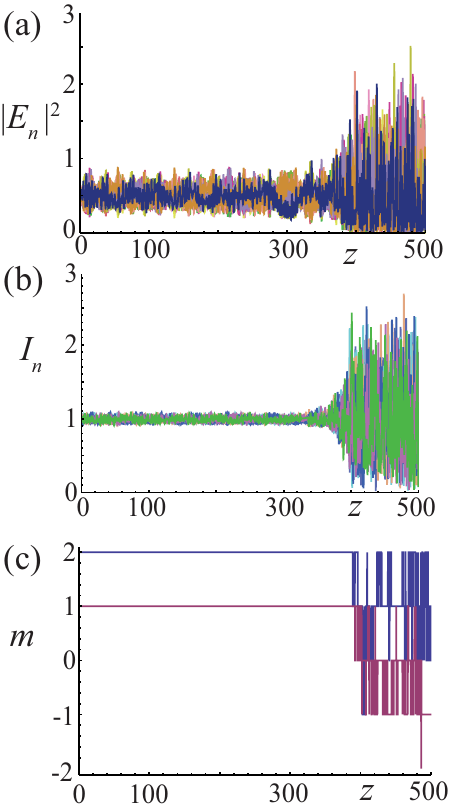}
\caption{The adiabatic dynamics of the degenerate mixed- vortex family, with
$P=1,\protect\phi =\protect\pi /4$, $\protect\varphi =0$, topological
charges (2,1) and random perturbations added to the initial conditions at
the 5\% level. (a) On-site powers of both components. (b) The total power at
each site, $I_{n}=|A_{n}|^{2}+|B_{n}|^{2}$. (c) The topological charges.}
\label{fig:dynamics}
\end{figure}

In contrast to the picture for the HV model outlined above, instability can
occur during the adiabatic cycling of the mixed vortex modes, since their
stability  depends on $\varphi $. This is shown in figure \ref{fig:dynamics},
in which the initial condition is the perturbed $\varphi =0$ mode.
Initially, it experiences large oscillations in terms of its azimuthal
modulation, while preserving its vortex charges and the power sums $I_{n}$.
At $z\approx 400$, an unstable value of $\varphi $ is reached and an
instability emerges, leading to irregular dynamics in which neither the
topological charges nor $I_{n}$ are conserved. If, instead, the $\varphi
=\pi /4$ mode is used as the initial condition, this instability sets in
immediately.

\begin{figure}[tbp]
\center
\includegraphics[width=60mm]{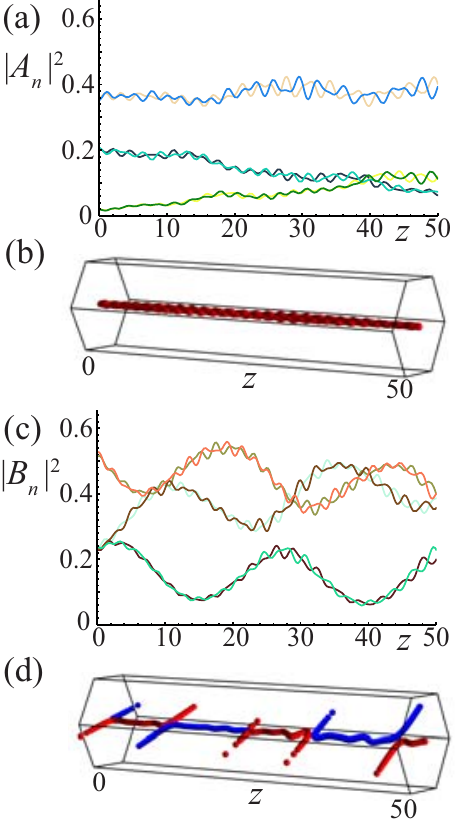}
\caption{The charge- flipping dynamics of the single component, with $P=1/2,%
\protect\phi =\protect\pi /5$, $\protect\varphi =\protect\pi /10$,
topological charges (2,-1), and random perturbations added to the initial
conditions at the 2\% level. (a,c): On-site powers for the two components.
(b,d): Vortex lines (red -- positive charge, blue -- negative charge).}
\label{fig:flip2}
\end{figure}

An additional feature demonstrated by the discrete system is the broken
conservation of the OAM. Namely, the two components can exchange their
angular momentum with the medium, as well as with each other. We show an
example of this in figure \ref{fig:flip2}, in which one component exchanges
the angular momentum with the medium, leading to the periodic reversal of
its topological charge, while the charge of the other component remains
conserved. This type of dynamics is generated by solutions (\ref{U})-(\ref%
{gam}) with $\phi \neq \pi /4$ and $\varphi $ chosen so as to make the value of $%
\gamma $ in (\ref{gam}) imaginary. The latter acts as a strong azimuthal
perturbation, leading to the charge flipping of the second component.

\section{Conclusions}

\label{sec:final}

We have studied the properties of composite vortex modes in circular arrays
of nonlinear waveguides. The stability hierarchy of discrete composite
vortices can be summarized by stating that the HV (hidden-vorticity) modes
suffer instabilities above a critical power, while explicit vortex modes
with high topological charges are stable. This hierarchy is opposite to that
in continua. Additionally, mixed-vortex modes with different topological
charges in the two components exist and can be stable. Degeneracies occur in
these families of composite vortex modes, which results in long-lived
breather states and persistent vortex-charge flipping.

It should be stressed that the analysis was performed here for the
relatively small ring chains, built of $N=6$ sites. For much larger rings,
one may expect a change in the stability and dynamics, as, for a fixed
diameter of the ring at $N\rightarrow \infty $, the system must go over into
the continuum limit, with its reverse picture of the stability domains for
the HV and explicit-vorticity modes.

These effects are visible at high powers required for the self-localization
in photonic lattices, therefore there is a possibility of observing them in
experiments similar to those that were aimed at studying discrete vortices
\cite{neshev2004, fleischer2004} and multivortex solitons \cite{terhalle2008}
in photorefractive crystals. A hexagonal lattice geometry corresponds to the
ring built of $6$ sites in our model. We can obtain appropriate experimental parameters for the observation of composite vortices from Ref. \cite{terhalle2009}, which studied double-charge discrete vortex solitons, corresponding to the scalar ($\phi = 0$) limit of our nonlinear modes. They used a 20mm long crystal with a bias voltage of 2.2kV/cm, a lattice wave beam with a total power of $75 \mu W$ and a $30\mu m$ period. Linear propagation was observed with a probe beam at $532nm$ with a total power of $P \approx 20nW$, while the nonlinear regime was reached at $P \approx 550 nW$. With these parameters, the propagation distance is long enough to observe the absence of discrete diffraction at high power (soliton formation) and the modulational instability of unstable modes. To observe composite vortex solitons, all that is required is to split the probe beam into two incoherent components, with the intensity and phase profiles generated using a spatial light modulator. Alternatively, our model can be realized directly in an integrated-optics
setting, using a femtosecond-laser written ring of nonlinear waveguides and two
incoherent beams \cite{femto}, although because of the weaker nonlinearity, significantly higher beam powers would be required.

\section*{Acknowledgment}

This work is supported by the Australian Research Council.

\end{document}